\documentclass{PoS}
\newcommand{\ig}{\includegraphics}

\title{Nucleon, Delta and Omega excited state spectra at three pion mass values}

\ShortTitle{Nucleon, Delta and Omega excited state spectra}

\author{John~ Bulava \\
NIC, DESY, Platanenallee 6, D-15738, Zeuthen, Germany \\
Email:  \email{john.bulava@desy.de}} 
\author{Robert~G.~Edwards, B\'{a}lint\ Jo\'{o}, David~G.~Richards \\
Thomas Jefferson National Accelerator Facility, Newport News, VA 23606, USA \\
Email: \email{edwards@jlab.org}, \email{bjoo@jlab.org}, \email{dgr@jlab.org}} 
\author{Eric~Engelson \\
In-Depth Engineering Corp., 11350 Random Hills Road, Fairfax, VA 22030, USA \\
Email: \email{engelson@gmail.com} }
\author{Huey-Wen Lin  \\
Department of Physics, University of Washington, Seattle, WA 98195, USA\\
Email: \email{hwlin@phys.washington.edu}} 
\author{Colin~ Morningstar \\
Department of Physics, Carnegie Mellon University,Pittsburgh, PA 15213, USA \\
Email: \email{colin\_morningstar@cmu.edu} }
\author{\speaker{Stephen J. Wallace} \\
        Department of Physics, University of Maryland, College Park, MD 20742, USA\\
        E-mail: \email{stevewal@umd.edu}}

\abstract{
The energies of the excited states of the Nucleon, $\Delta$
and $\Omega$ are computed in lattice QCD, using two light quarks and one strange quark
on anisotropic lattices.
The calculations are performed at three values  of the pion mass:
 $m_{\pi}$ = 392(4), 438(3) and 521(3) MeV.
 We employ the variational method with a basis of about ten interpolating operators enabling
 six energies  to be distinguished clearly in each irreducible representation of the octahedral group.
  We compare our calculations of nucleon excited states with the low-lying experimental spectrum.
  There is reasonable agreement for the pattern of states.  
}

\FullConference{The XXVIII International Symposium on Lattice Field Theory\\
                 June 14-19, 2010\\
                 Villasimius, Sardinia Italy}

\begin{document}

\section{Introduction}
The goal of determining the spectrum of hadron masses
from lattice QCD is addressed in this work by calculations of the excited state spectrum of the nucleon,
$\Delta$ and $\Omega$ using anisotropic lattices.\cite{Bulava:2010} 
Earlier excited-baryon analyses were based on quenched QCD~\cite{Basak:2007} and two-light-flavor ($N_f = 2$) QCD~\cite{Bulava:2009}.  In this work  we use ensembles of gauge configurations
developed in Ref.~\cite{Lin:2009} for $N_f=2+1$ QCD with two dynamical light quarks 
and one strange quark.  The lattices are $16^3\times 128$ with spatial and temporal lattice 
spacings $a_s $= 0.122 fm and $a_t$ = 0.035 fm.  

For families of particles with given isospin and strangeness, spectra are calculated
in the six double-valued irreducible representations (irreps) of
the octahedral group.  There are three irreps for
even-parity that are labeled with a $g$ subscript ({\it gerade})
and three for odd-parity that are labeled with a $u$ subscript
({\it ungerade}). They are: $G_{1g}, H_g, G_{2g}, G_{1u}, H_u$ and
$G_{2u}$. 

  Sets of seven to eleven three-quark
operators are used in each irrep and the variational method~\cite{Michael:1985, Luscher:1990} 
is used to extract energies of six states.  Most operators incorporate gauge-covariant displacements of the quarks relative to one another in order to obtain nontrivial shapes.~\cite{Basak:2005}   The recently developed ``distillation'' method~\cite{Peardon:2009} is used for quark smearing.  We start with 
a large set of operators in each irrep and then ``prune'' them to sets of about 10 that have
the lowest condition numbers.  That yields sets of approximately linearly-independent
operators that are suitable for calculations based on diagonalizing a matrix of correlation functions.  

\section{Results}
A detailed presentation of all of our results is given in Ref.~\cite{Bulava:2010}.  Here we present selected results for the nucleon and $\Delta$ excited states.

\begin{figure*}[h]
\ig[bb= 100 200 400 750]{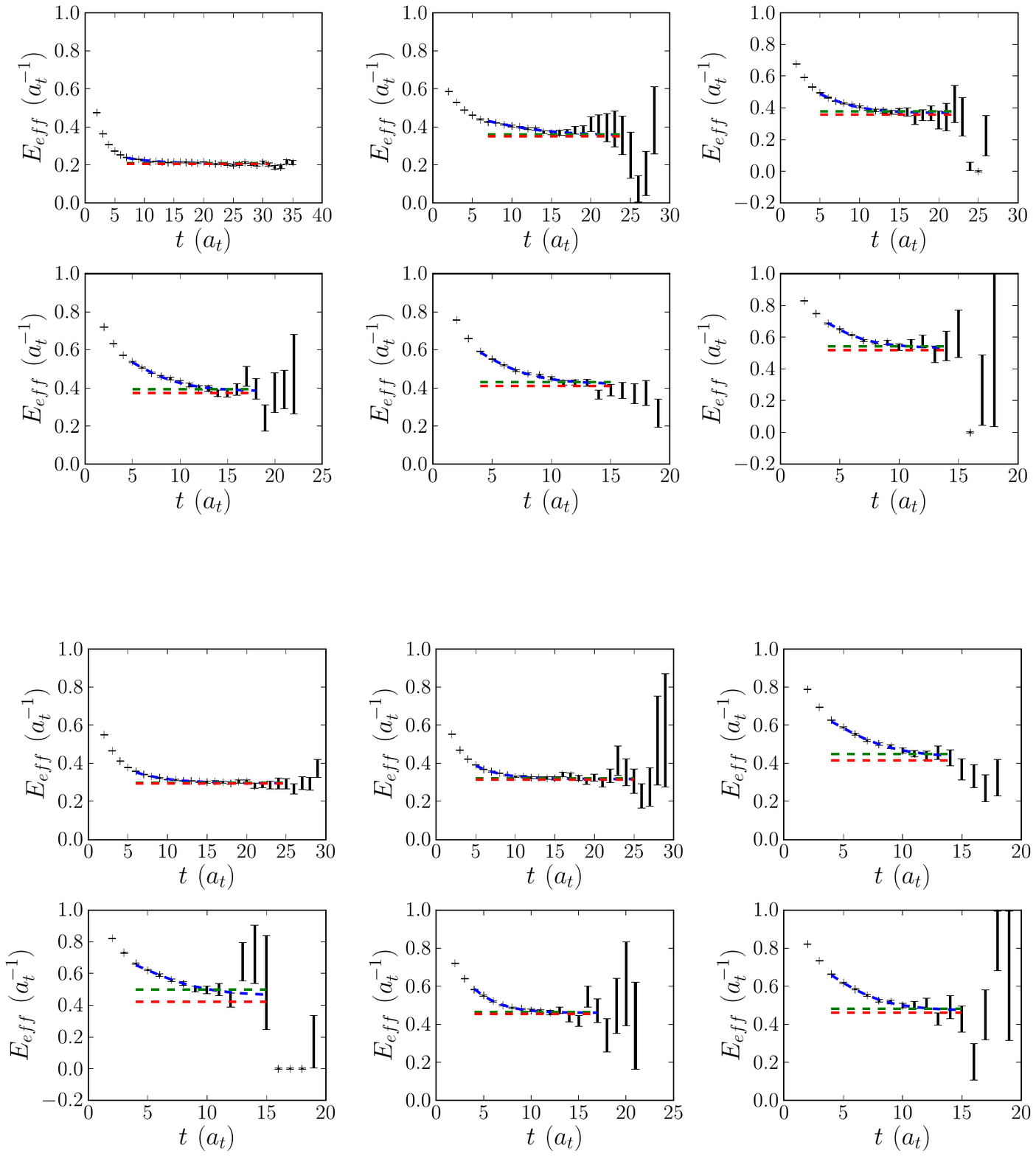}
\caption{ Nucleon $G_{1g}$ effective energies are shown for the lowest states in the upper
six graphs.  The effective energy increases from left to right along the first row and continues to increase from left to right along the second row.  The lower six graphs show
 nucleon $G_{1u}$ effective energies increasing in the same pattern.
Calculations are for $m_{\pi}= 392(4)$ MeV. Vertical bars show the effective energy and the curved dashed line shows the effective energy calculated from the fit function.  Horizontal dashed lines show the
fit results for $E \pm\sigma$ and their extent shows the fitting interval $(t_i,t_f)$.
}
\label{fig:meffplots-G1} 
\end{figure*}

Plots of the nucleon effective energies, calculated as 
\begin{equation}
E_{\rm eff}(t) = \frac{1}{2} ln\left( \frac{\widetilde{\lambda}(t-1)}{\widetilde{\lambda}(t+1)}\right),
\label{eq:Eeff}
\end{equation}
where $\widetilde{\lambda}(t)$ is an eigenvalue of the generalized eigenvalue problem,
are shown in Figure~\ref{fig:meffplots-G1} for the $G_{1g}$ and $G_{1u}$ irreps.  These plots show 
the values of $E_{\rm eff}$ obtained from Eq.~(\ref{eq:Eeff}) as vertical bars and $E_{\rm eff}$ calculated using the fit function,
\begin{equation}
\lambda_{fit}(t) = (1-A) e^{-E(t-t_0)} + A e^{-E^{\prime}(t-t_0)},
\label{eq:Cfit}
\end{equation}
 in place of $\widetilde{\lambda}(t)$ in Eq.~(\ref{eq:Eeff}) as curved dashed lines.  Comparison of the curved dashed lines with the bars from the lattice ensembles
shows the usefulness of two-exponential fits.  The term $Ae^{-E^{\prime}(t-t_0)}$
models the contributions of higher energy states at early times allowing the exponential term $(1-A)e^{-E(t-t_0)}$ to be determined over a larger fit window $(t_i, t_f)$ than would be possible using a single exponential.   Fit energy $E$ and uncertainty of the fit energy, $\sigma$, are shown by dashed horizontal lines
 at $E +\sigma$ and $E-\sigma$ extending over the fit window.
 Note that the statistics allow credible determinations of six energy levels in each irrep.  This
provides evidence that quark smearing based on ``distillation'' is effective with regard to suppressing  
 high-frequency fluctuations in the gauge ensembles. 

\begin{figure*}[t]
\hspace{-1.6in}
\ig[bb= -30 20 300 711]{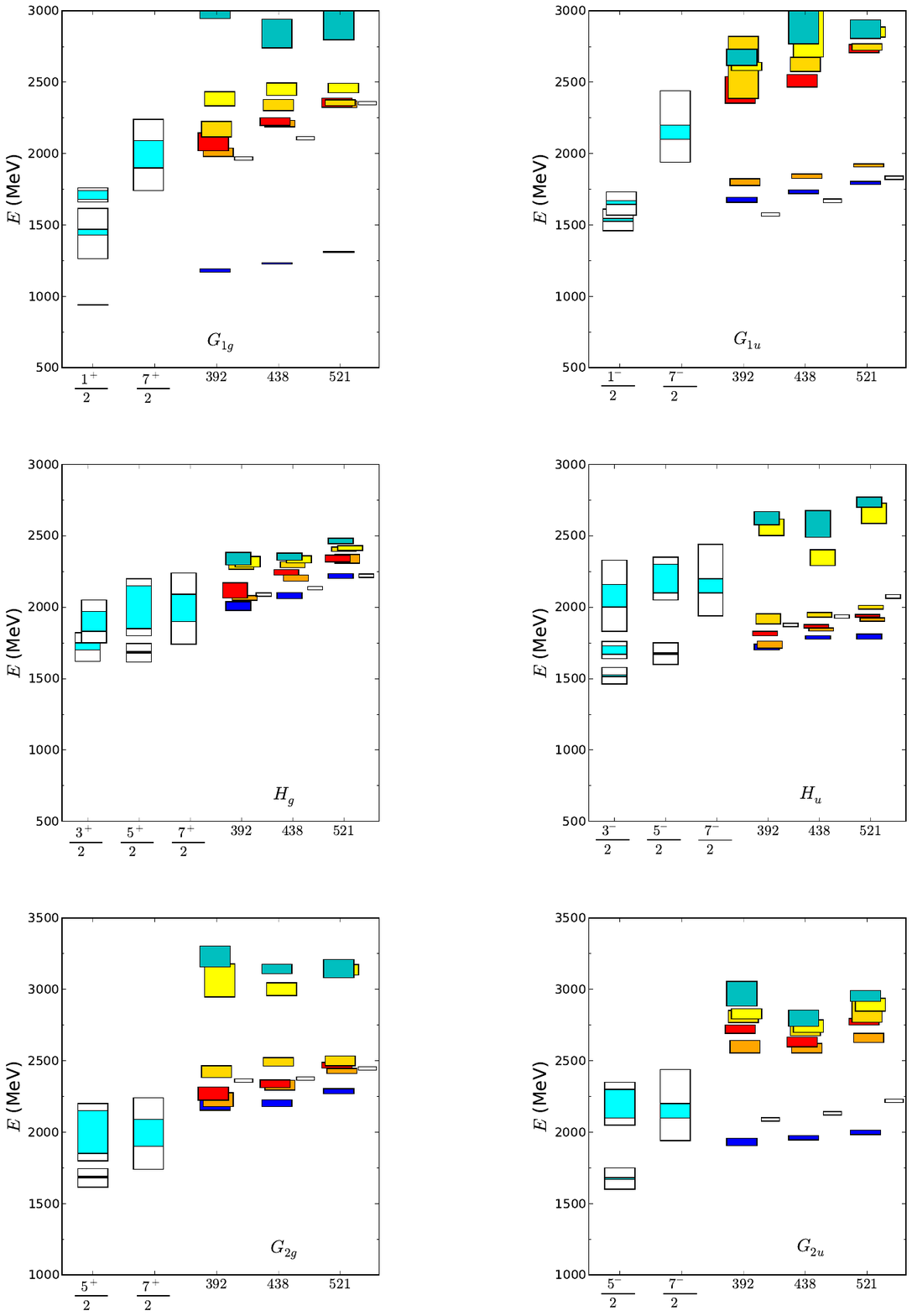}
\vspace{-2in}\caption{Spectra for isospin $\frac{1}{2}$ (nucleon family) at three values of $m_{\pi}$ 
in each irrep of the cubic group are compared with experimental spectra.   Columns labeled 
by $m_{\pi}$ =  392, 438 and 521 MeV show lattice spectra.  The 
$G_{1g}$ and $G_{1u}$ spectra in the $m_{\pi}$ = 392 MeV
column are obtained from the plots of
Fig. 1.  Boxes extend from $E-\sigma$ to $E+\sigma$.  Two, three and four-star experimental 
resonances are shown to the left of lattice spectra in columns labeled by their $J^P$ values.  Each $J^P$ value listed has a 
subduction to the lattice irrep shown.
Each box for an experimental resonance has height equal to the full decay width and an inner box (color aqua) showing the uncertainty in the Breit-Wigner energy. Open boxes show the thresholds for 
multiparticle states.  }
\label{fig:G1gHgG2gG1uHuG2u-nucleon-boxplots}
\end{figure*}

\begin{figure*}[t]
\hspace{-1.6in}
\ig[bb= -30 20 300 711]{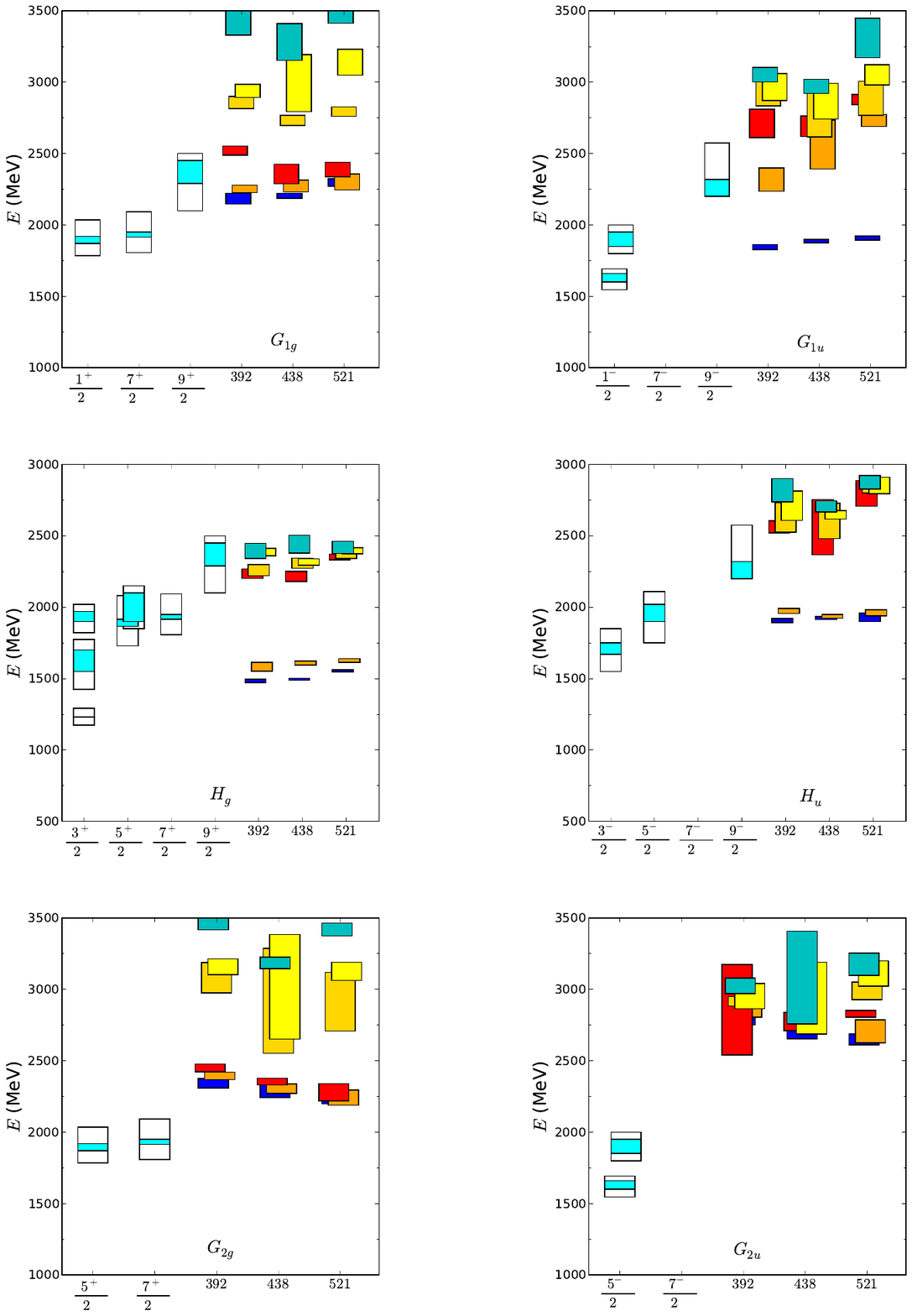}
\vspace{-2in}\caption{Spectra for isospin $\frac{3}{2}$ ($\Delta$ family) at three values of $m_{\pi}$ 
in each irrep of the cubic group are compared with experimental spectra.   Columns labeled 
by $m_{\pi}$ =  392, 438 and 521 MeV show lattice spectra.    Boxes extend from $E-\sigma$ to $E+\sigma$.  Two, three and four-star experimental 
resonances are shown to the left of lattice spectra in columns labeled by their $J^P$ values.  Each $J^P$ value listed has a 
subduction to the lattice irrep shown.
Each box for an experimental resonance has height equal to the full decay width and an inner box (color aqua) showing the uncertainty in the Breit-Wigner energy.  }
\label{fig:G1gHgG2gG1uHuG2u-delta-boxplots}
\end{figure*}

The energies obtained from the $G_{1g}$ and $G_{1u}$ effective mass plots of Fig.~\ref{fig:meffplots-G1}  are shown as boxes extending from $E-\sigma$ to $E+\sigma$ in Fig.~\ref{fig:G1gHgG2gG1uHuG2u-nucleon-boxplots}.  We
show nucleon energies that are obtained in the same manner as shown in 
Fig.~\ref{fig:meffplots-G1} for all irreps of the octahedral group and three pion masses. 
Experimental spectra are shown to the left of lattice energies for the spins and parities that have subductions to the lattice irreps.
  Lattice and experimental spectra are shown in Fig.~\ref{fig:G1gHgG2gG1uHuG2u-delta-boxplots} for the $\Delta$ family.  See Ref.~\cite{Bulava:2010} for the $\Omega$ spectra. 
  
  In the nucleon spectra, there is good evidence for a spin $\frac{5}{2}^-$ state. 
  We find nearly degenerate $H_u$ and $G_{2u}$ partner states, which is the signature of 
  spin $\frac{5}{2}^-$.   However, other spins are difficult to identify because there are many 
  nearly-degenerate states, within uncertainties.  It is a near-term goal within the 
  collaboration to address spin identification by using operators that are subduced from 
  continuum spins. 
  
Our lattice spectra show scant evidence for multiparticle states even though many energies lie above
the relevant thresholds. This is probably because single-hadron operators are used. There is an
 inference for a multiparticle contribution in that we find four low-lying states in $H_u$  while there are
 three low-lying experimental states that have subductions to $H_u$.  There is a threshold for a multiparticle state in the same energy range.  Our lattice results agree with the experimental
pattern if one of the four low-lying $H_u$ states is multiparticle.  However, we cannot identify the 
multiparticle states in the spectrum.  It is a near-term goal within the collaboration to incorporate 
multiparticle operators that couple directly to such states.  

Some lattice states appear to be ``squeezed'' by the small lattice volume used. They show up at higher energies than would be the case in a larger volume.   The $G_2$ states require partner states in
other irreps, such as $H$, in order to realize all the magnetic substates for a given spin.  The partners should be close to the same energy.  However, in the $\Delta$ spectra of Fig. 3 we find $G_2$ states at high energies without suitable partners being evident.  Possibly they have been ``squeezed''. 
It is also a goal to perform calculations of spectra at larger volumes.  

Although we do not attempt to extrapolate energies to $m_{\pi}$ = 140 MeV, 
it is evident from Figs. 2 and 3 that the lowest-energy states on the lattice tend toward the
energies of the physical resonances as the pion mass decreases.  Decreasing the pion mass 
is an obvious goal but we recognize that it entails a more complex analysis for excited states that can decay.

%

\section{Summary}

This work represents a milestone in our long-term research program aimed at
   determining the spectra of baryons in QCD.  It provides the first spectrum for
   N, $\Delta$ and $\Omega$ baryons based on $N_f=2+1$ QCD with high
   statistics.   A large number of baryon operators is used to calculate
   matrices of correlation functions.  They are analyzed using the variational method
   with fixed eigenvectors.
   The analysis provides spectra at three pion masses:
   $m_{\pi}$ =  392(4) MeV, 438(3) MeV and 521(3) MeV.

   The lattice volume and pion masses used give considerably higher
   energies than the experimental resonance energies.  However, there is reasonable agreement of the overall
   pattern of lattice and experimental states.  One exception is that almost all $\Delta$ states
   in the $G_2$ irrep are too high.  That may be caused by a volume that is too small for
   highly excited states.

\acknowledgments

This work was done using the Chroma software suite~\cite{Edwards:2005} on clusters at Jefferson Laboratory and the Fermi National Accelerator Laboratory using time awarded under the USQCD Initiative.
JB and CM acknowledge support from U.S. National Science Foundation Award
PHY-0653315.  EE and SW acknowledge support from U.S. Department of Energy contract
DE-FG02-93ER-40762.  HL acknowledges support from U.S. Department of Energy contract
DE-FG03-97ER4014.  BJ, RE and DR acknowledge support from U.S. Department of Energy contract DE-AC05-060R23177, under which Jefferson Science Associates, LLC, manages and operates Jefferson Laboratory.  BJ and RE acknowledge support under U.S. Dept. of Energy SciDAC contracts DE-FC02-06ER41440 and DE-FC02-06ER41449.  The U.S. Government retains a non-exclusive, paid-up, irrevocable,
world-wide license to publish or reproduce this manuscript for U.S. Government purposes.

\end{document}